%% file: main.tex
\pdfoutput=1
\documentclass{article}
\usepackage{fullpage}
\usepackage{graphicx}
\usepackage{amsmath, amssymb, amsthm}
\usepackage{thm-restate}
\usepackage{algorithm}
\usepackage[noend]{algpseudocode}
\usepackage{tikz}
\usepackage{subcaption}
\usetikzlibrary{arrows.meta,calc,positioning}
\usepackage[colorlinks=true,allcolors=blue]{hyperref}

\newtheorem{lemma}{Lemma}
\newtheorem{theorem}{Theorem}
\newtheorem{definition}{Definition}

\newtheorem{corollary}{Corollary}
\newtheorem{assumption}{Assumption}

\newcommand{\tmix}{\tau_{mix}}

\usepackage{todonotes}
\definecolor{xred}{HTML}{F19F9E}

\title{Expander Routing under Adversarial Deletions via Temporal Aggregation}
\author{Bernhard Haeupler\thanks{Partially funded by
the Ministry of Education and Science of Bulgaria's support for INSAIT as part of the Bulgarian National Roadmap
for Research Infrastructure and through the European Research Council (ERC) under the European Union's Horizon
2020 research and innovation program (ERC grant agreement 949272).}\\
\normalsize INSAIT, Sofia University ``St. Kliment Ohridski'' \& ETH Zurich
\and
Anton Paramonov\\
\normalsize ETH Zurich\\
}
\date{}

\begin{document}
	
	\maketitle
	
	\begin{abstract}
		In expander routing, we seek to establish paths in the expander graph between given source-target pairs 
        while minimizing both \emph{dilation} (maximum path length) and \emph{congestion} (maximum edge load). 
        We study this problem in a decremental dynamic setting where the underlying graph remains an expander 
        despite ongoing edge deletions, and source-target pairs can be added or removed. Our goal is to maintain paths under these updates while keeping dilation and congestion low and performing low computation per update.
        
        We present a randomized algorithm that maintains paths with polylogarithmic dilation and congestion 
        against an adaptive adversary, while incurring only polylogarithmic amortized computation per update, 
        all with high probability. Our approach leverages a recent technique called Temporal Aggregation to 
        maintain a collection of low-congestion, low-dilation random walks resilient to adaptive adversaries.
	\end{abstract}

	\section{Introduction}
	In the \emph{routing problem}, we are given a graph and pairs of nodes in this graph, called demands, between which we need to establish paths. The challenge is to come up with paths that are short but also have low congestion, i.e., the maximal number of paths passing through a given edge.
	
	The latter is not always achievable. For example, a node may participate in far more demands than its degree allows, or there may be a small cut separating many demand endpoints. This motivates restricting attention to graph classes in which the cut obstruction does not arise once the per-vertex feasibility condition holds. In particular, one can study graphs for which, whenever every node $v$ is the source of at most $\deg(v)$ demands, there is no sparse cut separating many demand endpoints. Such graphs are (essentially) \emph{expanders}, and the resulting restriction of the routing problem is known as \emph{expander routing}. Remarkably, these necessary conditions are also sufficient: expander routing always admits a solution with polylogarithmic congestion and path length. Moreover, in the \emph{static} setting---where the graph and the demand set are fixed---such a solution can be computed in polynomial time~\cite{chang2024deterministic}.
	
	Practical systems, however, are rarely static. Demands typically arrive and depart over time, and the underlying network itself may change. The challenge is even sharper under an \emph{adaptive adversary} that observes the algorithm's current state and chooses updates intended to force failure. A trivial approach is to recompute a full routing after every update, but this can incur large and unnecessary overhead. This leads to the following question:
	
	\begin{quote}
		\emph{Can we maintain expander routing efficiently under adaptive adversarial updates?}
	\end{quote}
	
	We answer this in the affirmative. Concretely, we give a maintenance scheme that preserves polylogarithmic congestion and path length while requiring only polylogarithmic amortized computation per adversarial update.
	
	\subsection{Dynamic Expander Routing Problem}
	We now formally define the \emph{dynamic expander routing} problem. Initially, the algorithm is given a graph $G_1$ on $n$ vertices and a set of demands $\mathcal{D}_1 \subseteq V(G_1)\times V(G_1)$. The algorithm may perform a polynomial-time setup phase, after which it outputs a collection of paths $\mathcal{P}_1$ in $G_1$ such that every demand $(s, t) \in \mathcal{D}_1$ is served by some path $p \in \mathcal{P}_1$, meaning $p$ connects $s$ and $t$.
	
	The process evolves over $T = \mathrm{poly}(n)$ time steps. At each time step $1 < t \leq T$, an adaptive adversary---having observed the current maintained paths---modifies the instance. Specifically, it deletes some edges from $G_{t-1}$ to obtain a new graph $G_t$, and it may also add or delete demands from $\mathcal{D}_{t-1}$ to obtain a new demand set $\mathcal{D}_t$. After these changes, the algorithm must update its solution from $\mathcal{P}_{t-1}$ to $\mathcal{P}_t$ so that every demand in $\mathcal{D}_t$ is served.
	
	The goal is to maintain a set of paths $\mathcal{P}_t$ with small dilation, i.e., the maximal path length, and small congestion, i.e., the maximum number of paths passing through any edge, at all times, while also minimizing the computation performed after the setup phase.
	
	\paragraph*{Adversary constraints.} 
	We impose three constraints on the adversary. First, we assume a feasibility condition on demand endpoints: 
	
	\begin{assumption}
		\label{assump:vertex feasibility}
		For every time $t \in [T]$ and every node $u \in V(G_t)$, the number of demands in $\mathcal{D}_t$ that include $u$ is at most the degree of $u$ in $G_t$.
	\end{assumption}
	
	Without this assumption, the degree of a single node becomes an unavoidable bottleneck, and low-congestion routing cannot be guaranteed.
	
	Second, we require that the graph remains an expander.
	
	\begin{assumption}
		\label{assump:always expander}
		There exists an integer $\tmix$ such that for every point in time $t \in [T]$, the mixing time of $G_t$ is no more than $\tmix$.
	\end{assumption}
	
	This assumption can be ensured via \emph{expander pruning} \cite{pruning}: after the adversary deletes edges, a pruning procedure may delete additional edges to restore expansion. Notably, if the adversary deletes $q$ edges in total, then pruning deletes $\widetilde{O}(q)$ edges and performs $\widetilde{O}(q)$ computation in total.
	
	Once we allow deleting $\widetilde{O}(q)$ edges in addition to adversary, we can as well assume that $q \leq \frac{|E(G_1)|}{\log^cn}$, for some constant $c$, since once $q$ reaches this threshold, we can delete all $|E(G_1)|$ edges and leave the graph empty.
	
	\begin{assumption}
		\label{assump:sublinear edge deletions}
		The total number of adversarial edge deletions is no more than $\frac{|E(G_1)|}{\log^cn}$ for some constant $c$.
	\end{assumption}
	
	Concretely, it suffices to take $c = c_0 + 1$, where $c_0$ is the constant such that, after $q$ deletions, the number of touched hypercube edges is at most $q\log^{c_0} n$ (Lemma~\ref{lem:few touched edges}); see the proof of Lemma~\ref{lem:path success}.

	\subsection{Dynamic Hypercube Embedding Problem}
	We show that dynamic expander routing can be reduced to a structured special case, namely \emph{dynamic generalized hypercube embedding}. A generalized hypercube is a graph that retains the hierarchical, bit-fixing-like connectivity of a hypercube, but is defined for an arbitrary number of vertices (not necessarily a power of two).
	
	\begin{definition}
		A \emph{generalized hypercube} is a graph on $n$ vertices labeled $1,2,\dots,n$. For each $u\in[n]$, the edge set contains the edges
		$(u,u+1),\ (u,u+2),\ (u,u+4),\ \ldots,\ (u,u+2^i)$,
		where $i$ is the largest integer such that $u+2^i\le n$.
	\end{definition}
	For convenience, in this work, we sometimes refer to a generalized hypercube simply as a hypercube.
	
	In the dynamic hypercube embedding problem, the algorithm is presented with a sequence of undirected graphs $G_1,G_2,\dots,G_T$, revealed online. The sequence is chosen by an adaptive adversary: each graph $G_{t}$ is obtained from $G_{t-1}$ by deleting a subset of edges after observing the algorithm's behavior on $G_{t-1}$. However, for every $t\in[T]$, the graph $G_t$ is guaranteed to be a good expander, that is, its mixing time satisfies $\tmix(G_t)\le \tmix$, where $\tmix=\mathrm{polylog}(n)$ is a known global bound.
	
	At each step $t > 1$, the algorithm must construct an embedding of a generalized hypercube $H_t$ onto $G_t$. For $t = 1$, we assume that the embedding is given. The embedding consists of a bijection $\xi_t:V(H_t)\to E(G_t)$ that assigns each hypercube vertex to a unique edge of $G_t$. Further, for every hypercube edge $(u,v)\in E(H_t)$, the algorithm must specify a path in $G_t$ connecting $\xi_t(u)$ and $\xi_t(v)$.
	
	The objective is to minimize, over all time steps, the maximum congestion and maximum dilation of the maintained paths, as well as the total computation performed across the update sequence.
	
	\subsection{Our Results}
	Our main result gives an efficient solution to dynamic expander routing against an adaptive adversary, optimal up to polylogarithmic factors.
	
	\begin{restatable}{theorem}{dynamicExpanderRoutingThm}
		\label{thm:expander routing}
		Under Assumptions~\ref{assump:vertex feasibility}--\ref{assump:sublinear edge deletions} with $\tmix = \mathrm{polylog}(n)$, there exists a randomized algorithm that solves the dynamic expander routing problem in the presence of an adaptive adversary such that, with high probability, at any point in time, the collection of paths serving the current demands has dilation and congestion at most $\widetilde{O}(1)$. Moreover, if the adversary performs $q = \mathrm{poly}(n)$ modifications in total (edge deletions and demand insertions/deletions), then with high probability the algorithm performs $\widetilde{O}(q)$ total computation.
	\end{restatable}
	
	A key building block is the ability to maintain, dynamically, a low-congestion and low-dilation embedding of a generalized hypercube into the changing expander.
	
	\begin{restatable}{theorem}{hypercubeEmbeddingThm}
		\label{thm:hypercube embedding}
		Assume $\tmix = \mathrm{polylog}(n)$, the initial embedding has congestion and dilation $\widetilde{O}(1)$, and the total number of edge deletions satisfies Assumption~\ref{assump:sublinear edge deletions}. Then there exists a randomized algorithm that solves the dynamic hypercube embedding problem in the presence of an adaptive adversary such that, with high probability, at any point in time, the collection of paths corresponding to the edges of a hypercube has dilation and congestion at most $\widetilde{O}(1)$. Moreover, if the adversary performs $q$ edge deletions in total, then with high probability the algorithm performs $\widetilde{O}(q)$ total computation.
	\end{restatable}
	
	We now give a high-level overview of how we achieve these results.
	
	\subsection{Our Approach}
	\paragraph*{Reducing to routing in hypercube.} A central special case of static expander routing is routing in the hypercube, which admits a particularly simple scheme due to Valiant. To route between two vertices $s$ and $t$, one samples a uniformly random intermediate vertex $w$ and routes from $s$ to $t$ via $w$, using a bit-fixing path from $s$ to $w$ and from $w$ to $t$. Under this scheme, if each node participates in at most $O(\log n)$ demands, then the congestion and dilation are $O(\log n)$ whp.
	
	Our approach leverages this idea in a two-step manner. First, we maintain an embedding of a virtual generalized hypercube into the underlying expander. Second, we route demands in this virtual hypercube using a Valiant-like strategy, and then project the resulting hypercube route back to the original expander using the maintained embedding.
	
	To illustrate the reduction, consider serving a demand $(s,t)\in V(G)\times V(G)$. Choose arbitrary edges incident to $s$ and $t$, say $e_1$ and $e_2$. Using the embedding $\xi$, we translate these to hypercube vertices $\xi^{-1}(e_1)$ and $\xi^{-1}(e_2)$, route between them in the (virtual) hypercube, and then map each hypercube edge to its projection in $G$. This yields a valid route from $s$ to $t$ in the original expander.
	
	\begin{figure}[ht]
		\centering
		
		\begin{subfigure}[t]{0.5\textwidth}
			\centering
			\resizebox{\linewidth}{!}{\input{pics/hypercube_embedding1.tikz}}
			\caption{Establish a Valiant-like route in the hypercube between the pre-images of edges incident to $s$ and $t$.}
		\end{subfigure}\hfill
		\begin{subfigure}[t]{0.44\textwidth}
			\centering
			\resizebox{\linewidth}{!}{\input{pics/hypercube_embedding2.tikz}}
			\caption{Project the established route back into $G$ as a concatenation of the projections of the hypercube edges.}
		\end{subfigure}
		
		\caption{A two-step method for routing between $s$ and $t$ in $G$ via an embedded hypercube $H$.}
	\end{figure}
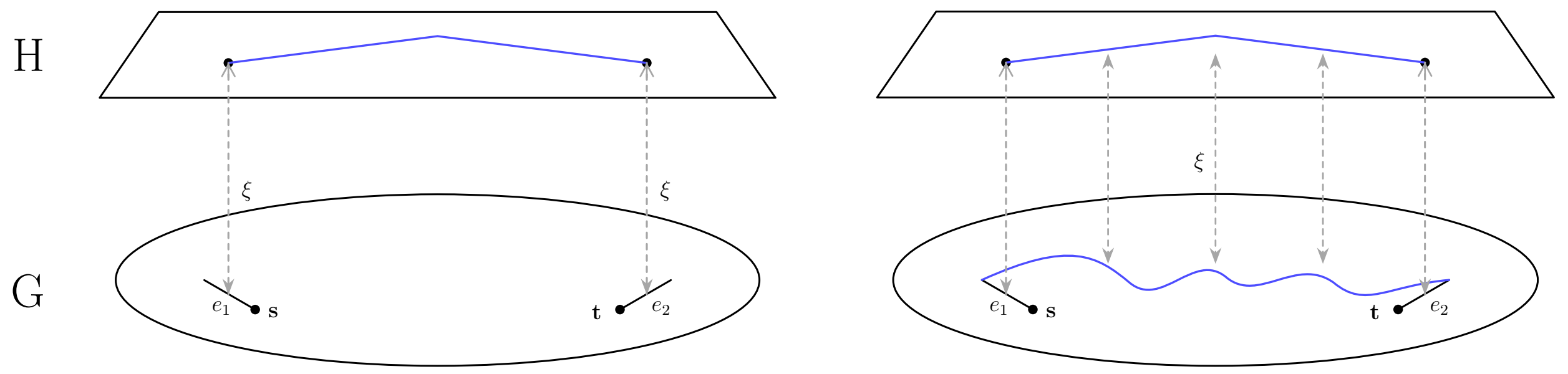
	
	\paragraph*{Using Temporal Aggregation.}
	Both in hypercube embedding and in hypercube routing, our constructions rely on random walks. In the static setting, random walks are an attractive tool: they naturally spread load and prevent any single edge from becoming overly congested. In the dynamic setting with an adaptive adversary, however, this robustness can break down. Since the adversary observes the algorithm’s current state, it can trigger repeated resampling of a walk---for example, by deleting an edge currently used by that walk---and thereby bias the maintained set of walks toward a chosen region of a graph, thus overcongesting it.
	
	One particularly damaging strategy is a \emph{resample-until-hit} attack, where the adversary repeatedly forces resampling until a chosen walk happens to traverse a targeted edge, and then repeats this process across many walks to accumulate congestion on that edge. To defend against such attacks, the algorithm must resample random walks proactively.
	
	To this end, we employ a recent framework known as \emph{Temporal Aggregation} \cite{temporalAggregation}. This scheme performs few proactive resamples, ensuring, however, regardless of the adversary’s strategy, that the collection of maintained random walks induces low congestion at every time step with high probability.
	
	\section{Related Work}
	
	\textbf{Static General Graph Routing.}
	The field of routing was hugely pioneered by R\"acke \cite{racke2002minimizing,racke2008optimal} whose endeavor culminated in designing a routing scheme for general graphs, that can produce a set of paths to serve the demands in polynomial time with congestion that is only a factor of $O(\log n)$ greater than the optimal congestion for the given problem instance. Furthermore, R\"acke's scheme is oblivious, meaning the algorithm's choice of path for a particular demand does not depend on the choices for other demands.
	
	However, R\"acke's results do not capture the dilation of paths. This gap was addressed in a work by Ghaffari, Haeupler, and Zuzic \cite{ghaffari2021hop}, which provides a polynomial time oblivious scheme that is $\widetilde{O}(1)$ competitive against the optimal $\mathrm{congestion} + \mathrm{dilation}$ for the problem instance.
	
	This result was further adapted to a CONGEST model \cite{haeupler2022hop} with an additional property of every node storing only a little additional information.
	
	\textbf{Static Special Case Routing.}
	The first work on oblivious routing is by Valiant and Brebner \cite{valiant1981universal}, who gave an elegant way to route demands in a hypercube with $O(\log n)$ dilation and congestion whp. Other special case results include studies on hole-free graphs \cite{busch2010optimal}, geometric graphs \cite{busch2005oblivious}, meshes \cite{busch2008optimal}, and other \cite{rabin1989efficient,scheideler2006universal,upfal1984efficient,racke2009survey}.
	
	\textbf{Static Expander Routing.}
	The expander routing problem was first studied by Ghaffari et al. \cite{originalExpanderRouting} who considered it in the CONGEST model. There, the demand $(u, v)$ means that $u$ wants to send a token to $v$, and the goal is to send all the tokens in as few rounds as possible. In \cite{originalExpanderRouting}, the round complexity of the problem was shown to be $\tmix(G)\cdot2^{O(\sqrt{\log n \cdot \log \log n})}$ with $\tmix(G)$ being a mixing time of the underlying graph $G$. This was further improved by \cite{newExpanderRouting}, which achieves $\tmix(G)\cdot 2^{O(\sqrt{\log n})}$ rounds. Algorithms in \cite{originalExpanderRouting} and \cite{newExpanderRouting} are randomized, which was addressed in works \cite{deterministicExpanderDecomposition,chang2024deterministic} that obtained a deterministic CONGEST algorithm for expander routing in $O(\mathrm{poly}(\phi^{-1}(G))\cdot2^{O(\sqrt{\log n \cdot \log \log n})})$ rounds with $\phi(G)$ being a conductance of the underlying graph $G$.
	
	Overall, expander routing has proven to be useful in many applications, such as distributed MST construction \cite{originalExpanderRouting}, subgraph finding \cite{trianglesThroughDecomposition, trianglesThroughDecomposition2}, and various data-summarization tasks \cite{su2019distributed}. Further, expander routing allows us to transform a large class of work-efficient PRAM algorithms into CONGEST algorithms with small overhead \cite{newExpanderRouting}, perform a smooth analysis for distributed MST \cite{chatterjee2020distributed}, and design sparsity-aware algorithms for various shortest path computation tasks in the CONGEST model \cite{censor2021sparsity}.

	\section{Preliminaries and Notation}
	We use $n$ as an upper bound on the number of vertices in $G_t$ for all $t$. One can assume $n := |V(G_1)|$. Similarly, $m$ is the upper bound on the number of edges in $G_t$. Since we assume $G$-s to be connected, we use $O(\log n)$ instead of $O(\log m )$ everywhere.  When the exact time step is not relevant, we omit specifying it and simply write $G$ and $H$. We use $\widetilde{O}$ notation to hide polylogarithmic factors. By with high probability (w.h.p.) we mean the probability of at least $1 - \frac{1}{n^c}$ for some $c \geq 1$.
	
	A random walk on a graph is a discrete process that starts at some vertex and, at each step, travels to a uniformly random neighbor of the current vertex. A lazy random walk is the one that, alternatively, with probability $1/2$, stays at the same vertex. It is known that the distribution of a lazy random walk converges to a stationary distribution in which each vertex $v \in G$ has a probability $\frac{\mathrm{deg}_G(v)}{2|E(G)|}$. Alternatively, we will sometimes speak about edge random walks, the difference being that the random walk starts from an edge and travels to an adjacent edge. The reason to shift to this perspective is that the edge random walk possesses a uniform stationary distribution over edges, which is convenient to work with. 
	
	The mixing time of $G$, denoted $\tmix(G)$, can be defined as the number of time steps for (any) initial distribution of a random walk on $G$ to get inverse-polynomially close to the stationary one in the sense of $L_2$ distance. With that, expanders can be defined as graphs having polylogarithmic mixing time. In some of our derivations, we implicitly assume $\tmix = \widetilde{O}(1)$.
	
	Throughout the paper, we will refer to classical Chernoff bounds. 
	
	Given an embedding $\xi$ of a hypercube $H$, we sometimes slightly abuse the notation and write $\xi(e)$ for $e \in E(H)$ to refer to the path in $G$ to which $e$ is mapped by the embedding. We call this path a projection of $e$ or a path corresponding to $e$, and we define a projection of a path in $H$ to be a concatenation of projections of edges constituting this path.  
	
	\paragraph*{Temporal Aggregation.}
	A key tool we will use repeatedly is \emph{Temporal Aggregation}. The motivation is the following. Our algorithms are randomized and rely on randomness for load balancing. In a static setting, where all random choices are made simultaneously, standard concentration arguments imply that the maximum load is small with high probability. In a dynamic setting with an \emph{adaptive} adversary, however, this guarantee can break down: if the adversary is allowed to repeatedly resample outcomes after observing them, it can bias the realized distribution and create excessive load.
	
	We illustrate this issue with two examples that are central to this paper. First, consider Valiant routing on the hypercube. If all demands are routed simultaneously, then with high probability, the maximum congestion is low. In the dynamic setting, an adversary that observes the current $u$--$v$ path can force it to be resampled, e.g., by deleting an edge on it or by deleting and re-inserting the demand $(u,v)$. It can then repeat this until the path traverses some fixed target edge $e$, and then proceed similarly for other pair $u',v'$, and so on. In polynomially many steps, this can force many routes to traverse $e$, thereby overcongesting it.
	
	As a second example, consider maintaining random walks in a graph that changes over time. Suppose that from each node we want to maintain $O(\log n)$ random walks of length $\ell$. If all these walks are sampled simultaneously, then with high probability the maximum congestion is $O(\ell \log n)$. But the same ``resample-until-hit'' attack applies here as well. Moreover, if the adversary can delete edges, it can increase the probability of certain random walks to hit a given edge.
	
	In both examples, it is natural for the algorithm to resample random paths from time to time in order to ``refresh'' the randomness and prevent the adversary from drifting too far from the intended distribution. However, fully resampling the entire system after every adversarial update would be prohibitively expensive. To address this, we use the \emph{Temporal Aggregation} framework of~\cite{temporalAggregation}, which was designed specifically for maintaining randomness in dynamic systems against adaptive adversaries.
	
	The framework in~\cite{temporalAggregation} is as follows. The system consists of $n$ objects and evolves in discrete time steps. At each time, an object corresponds to an element in some ground set, we call this element the realization of an object. At time $t$, after observing the state of the system at time $t-1$, the adversary specifies a distribution over the ground set for each object and resamples a subset of objects according to these distributions. Then, the algorithm may also resample a subset of objects of its choice, according to the specified distributions. Objects that are not resampled at time $t$ keep their previous realizations. The set of all objects' realizations at time $t$ is called a joint realization, and the set of all objects' realizations at time $t$ given they are all sampled at time $t$ is called a static realization.
	
	This formalism matches our needs. For instance, in the dynamic random-walk example, the objects correspond to individual random walks; their distributions are determined by the current graph; and objects' realizations come from a ground set of all paths of length $l$. The quantity we care about is the maximum congestion, which corresponds to the notion of \emph{load} in~\cite{temporalAggregation}. In our setting, the static realization has a load of $O(l\log n)$ whp, and we want to minimize the load of a joint realization.
	
	The paper~\cite{temporalAggregation} provides a resampling strategy called \emph{Greedy Temporal Aggregation}. Informally, if sampling all objects simultaneously would yield low load with high probability at each time, then Greedy Temporal Aggregation guarantees that the load remains low even in the dynamic, adversarially adaptive setting, at the cost of only a logarithmic-factor blow-up in both load and the number of resamples.
	
	\begin{theorem}[Theorem~2 of~\cite{temporalAggregation}]
		\label{thm:greedy temporal aggregation main}
		If the system evolves for $T$ time steps, and in each time step a static realization has load at most $L$ with probability at least $1-\varepsilon$, then Greedy Temporal Aggregation ensures that at time $T$ the joint realization has load at most $L\cdot O(\log n)$ with probability at least $1-\varepsilon\cdot T$.
		
		Moreover, if the adversary performs $q$ samples by time $T$, then Greedy Temporal Aggregation performs at most $q\cdot O(\log n)$ samples.
	\end{theorem}
	
	Theorem~\ref{thm:greedy temporal aggregation main} is stated for a fixed time $T$. Applying it at every time $t \le T$ and taking a union bound, the load bound holds simultaneously at all times $t\in[T]$ with probability at least $1-\varepsilon T^2$. Since in our applications $T = \mathrm{poly}(n)$ and $\varepsilon = n^{-c}$ for an arbitrarily large constant $c$, this is still a high-probability guarantee. Whenever we invoke the theorem ``at any point in time'' below, we mean this simultaneous version.
	
	To illustrate the power of this result, we show how this theorem allows us to extend static oblivious routing guarantees to a dynamic setting with an adaptive adversary essentially ``for free,'' up to logarithmic factors.
	
	\begin{definition}
		\label{def:oblivious routing}
		An \emph{oblivious routing scheme} is a randomized procedure that, given a graph $G$ and a single demand $(s,t)$, outputs an $s$--$t$ path in $G$ sampled from a distribution that depends only on $G$, $s$, and $t$. Each invocation uses fresh randomness, independent of all other invocations; in particular, the path chosen for one demand does not depend on the other demands or on the paths chosen for them. We say that the scheme yields congestion at most $L$ on $(G,\mathcal{D})$ with high probability if, when every demand in $\mathcal{D}$ is routed by an independent invocation, the maximum edge congestion is at most $L$ with high probability.
	\end{definition}
	
	\begin{corollary}
		\label{cor:temporal aggregation 4 routing}
		Consider a sequence of graphs $G_1,\ldots,G_T$ on $n$ vertices, where $T=\mathrm{poly}(n)$, and a sequence of demand sets $\mathcal{D}_1,\ldots,\mathcal{D}_T$. Suppose there exists an oblivious routing scheme such that for every $t\in[T]$, routing $\mathcal{D}_t$ in $G_t$ according to this scheme yields congestion at most $L$ with high probability.
		
		Then there exists an algorithm that maintains routing paths for $\mathcal{D}_t$ against an adaptive adversary such that the congestion stays $O(L\log n)$ at all times $t\in[T]$ whp. Moreover, if the adversary performs a total of $q$ edge deletions and demand insertions,  the algorithm performs at most $O(q\cdot L\log^2 n)$ invocations of the oblivious routing procedure whp.
	\end{corollary}
	\begin{proof}
		We map our setting to the framework of~\cite{temporalAggregation}. Each demand is an object. We say an object is adversarially resampled whenever the adversary either (i) introduces a new demand, or (ii) deletes an edge that the demand's current path uses. Resampling an object corresponds to recomputing its route using the oblivious routing scheme. The load at time $t$ is the maximum congestion of the current collection of paths at time $t$.
		
		Applying Greedy Temporal Aggregation, Theorem~\ref{thm:greedy temporal aggregation main} implies that at any time $t$, the maximum congestion is at most $O(L\log n)$ whp. In particular, when the adversary deletes an edge, at most $O(L\log n)$ demands are resampled whp. Therefore, over $q$ edge deletions, the adversary triggers at most $O(q\cdot L\log n)$ adversarial resamples whp. By Theorem~\ref{thm:greedy temporal aggregation main}, the algorithm performs only an additional $O(\log n)$ factor more resamples, for a total of $O(q\cdot L\log^2 n)$ oblivious-routing invocations whp.
	\end{proof}
	
	We want to remark that for our needs, a slight extension of the framework of \cite{temporalAggregation} is needed. Namely, \cite{temporalAggregation} assumes that the set of objects is static, whereas our applications require adding and removing objects. We show in the Appendix \ref{sec:appendix:extending ta} that this extension can be assumed without losing guarantees of Theorem \ref{thm:greedy temporal aggregation main}.

	\section{Dynamic Hypercube Embedding Algorithm}
	\label{sec:alg}

	In this section, we describe our algorithm that solves the Dynamic Hypercube Embedding problem optimally up to polylog factors. We start by defining the notion that resembles that of a bit-fixing path:
	
	\begin{definition}
		The \emph{greedy path} between nodes $u$ and $v$ in the generalized hypercube $H$ is defined as follows. If $u = v$, then the path is empty. Otherwise, w.l.o.g., assume that $u < v$. We pick the biggest $w \in V(H)$ s.t. $w \leq v$ and there exists an edge $(u, w) \in E(H)$, we add $(u, w)$ to the greedy path and construct the greedy path between $w$ and $v$ recursively.
	\end{definition}

	As described in the problem statement, the adversary constructs the next graph $G_{t+1}$ in the sequence by removing some edges from $G_t$, which corresponds to node deletions in $H_t$. This imposes two subtasks: (I) ensuring the graph maintained is a generalized hypercube despite some nodes being deleted, and (II) ensuring each edge of the maintained hypercube corresponds to an existing path in the underlying graph. 
	
	We now describe how our algorithm adapts $H_t$ and the embedding. Recall that nodes of the hypercube $H$ are associated with numbers $[1, \ldots, |V(H)|]$.
	
	\begin{algorithm}    
		\caption{Dynamic Hypercube Embedding. On $G$ Edges Deletion.}
		\label{alg:dynamic hypercube embedding}

		(I) Assume the vertex $u \in [|V(H_t)|]$ is deleted. If $u = |V(H_t)|$, we do nothing. Otherwise, we relabel the node $|V(H_t)|$ to $u$ but don't change its mapping by $\xi$. We then remove all the edges adjacent to deleted and relabeled nodes and \emph{establish} new edges $(u, u + 2^i)$ for all $i$ such that $u + 2^i \leq |V(H_{t+1})|$ and similarly $(u, u - 2^j)$ such that $u - 2^j \geq 1$.
		
		(II) When an edge is deleted in $G_t$, it might as well invalidate some paths that were traversing it. We therefore need to \emph{fix} all the edges from $H_t$ whose paths got invalidated by reestablishing those.
	\end{algorithm}
	
	\begin{definition}
		If an edge in $H$ is established or fixed at time $t > 1$, we call it \emph{new}. Otherwise, we call it \emph{original}.
	\end{definition}
	
	We now describe how to sample a path between edges $u$ and $v$ in $G$. We use this procedure for establishing paths for new edges.
	
	\begin{algorithm}
		
		\caption{Dynamic Hypercube Embedding. Sampling a $u-v$ path in $G$.}
		
		First, we run random walks from $u$ and $v$ of length $\tmix$, getting the endpoints $u' \in E(G)$ and $v' \in E(G)$ respectively. These endpoints correspond to $\xi^{-1}(u') \in V(H)$ and  $\xi^{-1}(v') \in V(H)$ respectively. Then, we sample a uniformly random node $w \in V(H)$ and construct greedy paths in $H$ between $\xi^{-1}(u')$ and $w$ and $w$ and $\xi^{-1}(v')$. If these greedy paths do not contain new edges nor edges used by some other greedy path, we establish a path between $u$ and $v$ as a concatenation of two sampled random walks and projections of two greedy paths. If either of the two greedy paths contains new edges or edges used by some other greedy path, we resample random walks and repeat this process until successful. 
	\end{algorithm}

	The final ingredient in our algorithm is the Temporal Aggregation of newly established or fixed edges.

	\begin{algorithm}		
		\caption{Dynamic Hypercube Embedding. Temporal Aggregation.}
		
		 We map our setting to the framework of \cite{temporalAggregation} as follows. Every edge $(u, v) \in E(H)$ of the hypercube corresponds to an object. The realization of the object $(u, v)$ at time $t$ is the path that corresponds to $(u, v)$ in $G_t$ between $\xi(u)$ and $\xi(v)$. Sampling an object means sampling a path as described above. Finally, we say that adversary adds an object $(u, v)$ at time $t$ if $(u, v)$ is a new edge added at time $t$, and we say that adversary samples an object $(u, v)$ if adversary deletes an edge in $G_t$ through which the path corresponding to $(u, v)$ was passing. In this setting, we employ Temporal Aggregation.
		
	\end{algorithm}

	It will be convenient for further analysis to introduce some definitions related to this algorithm.
	
	\begin{definition}
		If an edge in $H$ is either new or is used by some greedy path, we call it \emph{touched}.
	\end{definition}
	
	\begin{definition}
		The path between two nodes $u$ and $v$ in $G$ consists of two random walks and two greedy path projections. We call the random walks part the \emph{legs} of the $u-v$ path.
	\end{definition}

	\subsection{Proof Overview}
	Our goal is to show that the algorithm solves the problem optimally up to a polylogarithmic factor. Specifically, we aim to prove that all established paths have length $\widetilde{O}(1)$, that with high probability the congestion on every edge in $G$ remains $\widetilde{O}(1)$, and that the total number of computation steps is $\widetilde{O}(q)$ with high probability, where $q$ is the total number of adversarial edge deletions. In this section, we do the proof overview. We present formal statements in Section \ref{sec:formal statements}.
	
	\noindent\textbf{Short paths.} The path in $G$ between $u$ and $v$ consists of two random walks of length $\tmix$ and projections of two greedy paths.
	
	Since the greedy path in $H$ only passes through the untouched edges, whose paths were constructed at the initial embedding of the hypercube, the length of the path between $u$ and $v$ is at most $2\tmix + D_{init}\cdot L_{greedy}$, where $D_{init}$ is the maximal dilation of the initial embedding and $L_{greedy}$ is the maximal length of a greedy path in $H$. Lemma \ref{lem:greedy length} bounds $L_{greedy}$ as $O(\log n)$.
	
	\noindent\textbf{Low congestion.} In our analysis, we split the congestion introduced by paths into two parts: the congestion introduced by legs and the congestion introduced by the greedy path projections. The bound on the congestion from greedy path projections is straightforward. In particular, note that the number of greedy path projections passing through an edge $e \in E(G)$ is at most the number of the original edges whose projections pass through $e$, which is at most the maximal congestion of the original embedding. 
	
	To analyze the congestion coming from random walks, however, we refer to the guarantees of Temporal Aggregation, which yield the following crucial lemma:
	
	\begin{restatable}{lemma}{rwCongestion}
		\label{lem:rw congestion}
		With high probability, at any point in time $t$, for every edge $e \in E(G_t)$, the number of legs passing through $e$ is $\widetilde{O}(1)$.
	\end{restatable}

	Combined, these analyses result in the following lemma.
	\begin{restatable}{lemma}{lowCongest}
		\label{lem:low congest}
		With high probability, at any point in time, the maximal congestion of a hypercube embedding is $\widetilde{O}(1)$.
	\end{restatable}
	
	\noindent\textbf{Few computation steps.} The computation the algorithm performs can be analyzed per edge deletion. When an edge is deleted, the algorithm needs to (1) establish paths for new edges in $H$, (2) fix edges whose corresponding paths in $G$ were invalidated by an edge deletion, and (3) proactively resample edges through temporal aggregation. This way, bounding computation boils down to bounding the total number of paths to be (re)sampled and the computation needed to sample a single path. 
	
	\textit{Number of path (re)samples per edge deletion.} For one edge deletion, we establish $\mathrm{max\_degree}(H) = O(\log(n))$ new edges in $H$. Furthermore, Lemma \ref{lem:low congest} tells us that w.h.p., the number of edges to be fixed because of the edge deletion is $\widetilde{O}(1)$. Finally, Theorem \ref{thm:greedy temporal aggregation main} tells us that temporal aggregation only introduces an additional $O(\log n)$ factor to the number of edges to proactively resample compared to the number of newly established and fixed edges.
	
	\textit{Computation per path sampling.} Trying to sample a path once is cheap: the algorithm runs two random walks of length $\tmix$ in $G$ and computes two greedy paths in $H$, both straightforwardly in $\mathrm{polylog}(n)$ steps. However, the subtlety is that the obtained path might get rejected in case greedy paths go through touched edges, which will cause another attempt. Therefore, we need to upper bound the number of potential retries.  
	
	\begin{restatable}{lemma}{pathSuccess}
		\label{lem:path success}
		With high probability, at any point in time, the procedure of sampling a path succeeds after $O(\log n)$ tries.
	\end{restatable}
	
	Intuitively, the above lemma holds because there are few touched edges:
	\begin{restatable}{lemma}{fewTouchedEdges}
		\label{lem:few touched edges}
		After $q$ deletions, the number of touched edges is $\widetilde{O}(q)$. 
	\end{restatable}
	
	\subsection{Formal Statements}
	\label{sec:formal statements}
	In this section, we provide all the formal statements needed to prove that the Algorithm \ref{alg:dynamic hypercube embedding} results in low Dilation, Congestion, and Computation. 
	
	\subsubsection{Dilation}
	This subsection provides formal statements that show that the dilation of the paths that correspond to edges in $H$ is small at any time. 
	
	The following lemma bounds the length of greedy paths in $H$.
	\begin{lemma}
		\label{lem:greedy length}
		All greedy paths in $H$ are of length $O(\log n)$. 
	\end{lemma}
	\begin{proof}
		Consider a greedy path $u_1, \ldots, u_k$ between $u_1$ and $u_k$ with $u_1 < u_k$. Observe two things: First, $u_i < u_{i+1}$. Second, $u_{i + 1} - u_i \geq 2(u_{i + 2} - u_{i + 1})$. And since $u_k - u_1 = \mathrm{poly}(n)$, we deduce that $k = O(\log n )$.
	\end{proof}
	
	Using this lemma, we conclude that the maximal dilation at any time is polylogarithmic:
	\begin{lemma}
		\label{lem: hypercube embedding dilation}
		For any time $t$ and any edge $e \in E(H_t)$, the corresponding path to $e$ in $G_t$ has length $\widetilde{O}(1)$. 
	\end{lemma}
	\begin{proof}
		Consider an edge $(u, v)\in E(H_t)$. The path in $G_t$ between $\xi(u)$ and $\xi(v)$ consists of two legs of length $\tmix = \widetilde{O}(1)$ and two projections of the greedy paths. 
		
		Since the greedy path in $H$ only passes through the untouched edges, whose corresponding paths were constructed at the original embedding, the length of the path between $\xi(u)$ and $\xi(v)$ is at most $2\tmix + D_{orig}\cdot L_{greedy}$, where $D_{orig}$ is the maximal dilation of the original embedding, which is promised to be $\widetilde{O}(1)$, and $L_{greedy}$ is the maximal length of a greedy path in $H_t$, which Lemma \ref{lem:greedy length} bounds as $O(\log n)$.
	\end{proof}

	\subsubsection{Congestion}
	This subsection focuses on proving that with high probability, at any point in time, the maximal congestion is low. We analyze congestion that comes from greedy paths' projections and legs separately.
	
	The following lemma bounds the amount of congestion introduced by greedy paths' projections. 
	\begin{lemma}
		\label{lem:greedy per edge}
		At each time, at most $\widetilde{O}(1)$ greedy path projections pass through each edge in $G$. 
	\end{lemma}
	\begin{proof}
		By the algorithm, a greedy path in $H$ only takes untouched edges. This implies that if a projection of a greedy path goes through some edge $e \in E(G)$, then there exists an edge in $H$ on this greedy path which was established at time $1$ and was never fixed, and its corresponding path goes through $e$. But the number of such edges that are neither newly established nor were fixed and pass through $e$ is at most the maximal congestion of the original embedding, which is guaranteed to be $\widetilde{O}(1)$. Therefore, since by the algorithm at most one greedy path in $H$ can pass through such an edge, the number of greedy path projections traversing $e$ is at most $\widetilde{O}(1)$.
	\end{proof}
	
	\fewTouchedEdges*
	\begin{proof}
		Let us denote $X$, $Y$, and $Z$, respectively, the number of newly established edges, fixed edges that are not newly established, and edges traversed by a greedy path after $q$ edge deletions. Then by definition, the number of touched edges is $X + Y + Z$.
		
		For a single edge deletion, we establish $O(\log n )$ new edges, hence $X = O(q\cdot\log n)$. 
		
		Now, let us analyze how many edges in $H$ that were neither newly established nor fixed become fixed per edge deletion. Consider such an edge. The fact that it becomes fixed after an edge deletion implies that its corresponding path passed through a deleted edge. But then the number of such edges in $H$ is at most the congestion of an original embedding, which is $\widetilde{O}(1)$. Therefore, $Y = \widetilde{O}(q)$.
		
		Finally, each newly established or fixed edge introduces two greedy paths that touch every edge through which they go. However, Lemma \ref{lem:greedy length} stipulates that the length of a greedy path is $O(\log n)$, hence $Z = O((X + Y)\cdot\log n)$.  
	\end{proof}
	
	The lemma below intuitively implies that one touched edge disallows a relatively small number of greedy paths to be taken by the algorithm. 
	\begin{lemma}
		\label{lem:edge belongs greedy}
		Any edge $e\in E(H)$ belongs to at most $|V(H)|$ greedy paths in $H$.
	\end{lemma}
	\begin{proof}
		Define a \emph{stretch} of an edge $(u, v) \in E(H)$ to be $|u - v|$. Given that an edge belongs to a greedy path, this greedy path can be described by the stretches of the edges it contains. Note that for every greedy path, consecutive edges differ in stretch by at least a factor of $2$; furthermore, the stretches are monotonically decreasing. Therefore, since a stretch is at most $|V(H)|$ and at least $1$, an edge of stretch $l$ can potentially have at most $\log_2|V(H)| - \log_2l$ edges before it and $\log_2l$ edges after it in the greedy path. Which leaves us with $2^{\log_2|V(H)| - \log_2l}\cdot2^{\log_2l} = |V(H)|$ options.
	\end{proof}
	
	Next is the lemma that bounds the number of retries before a successful sample of a path.
	\pathSuccess*
	\begin{proof}
		Lemma \ref{lem:few touched edges} bounds the number of touched edges after $q$ edge deletions to be no more than $\log^{c_0}n\cdot q$ for some constant $c_0$. Leveraging Assumption \ref{assump:sublinear edge deletions} with $c = c_0 + 1$, the maximal number of edge deletions is $q:= \frac{m}{\log^{c_0 + 1}n}$, therefore ensuring that the number of touched edges is no more than $\frac{m}{\log n}$ at any point in time. Furthermore, since the number of deleted edges in $G$ is at most $q$, the number of vertices in $H$ at any point in time is at least $m - q = \Omega(m)$; therefore, the number of greedy paths in $H$ at any time $t$, denoted with $GP_t$, is $\Omega(m^2)$.
		
		Now, Lemma \ref{lem:edge belongs greedy} tells us that one edge belongs to at most $m$ greedy paths in $H$, hence, at any time, the total number of greedy paths that do not contain a touched edge is $GP_t - m \cdot\frac{m}{\log n }$. 
		
		Hence, since one endpoint of a greedy path is sampled almost uniformly and the other one purely uniformly, we conclude that sampling a greedy path with no touched edges is of probability $\frac{GP_t - m \cdot\frac{m}{\log n }}{GP_t} \pm\frac{1}{\mathrm{poly(n)}} = 1  - O(\frac{1}{\log n}) > \frac{1}{2}$, thus, by standard Chernoff bounds, after $O(\log n)$ tries, the algorithm will succeed with high probability. 
	\end{proof}
	
	The latter lemma allows us to show that although the distribution of paths' legs is different from the distribution of sample-once random walks due to their resample-until-success nature, those still induce a congestion comparable to the one induced by sample-once random walks. In particular, we get the following guarantee in the static setting.
	
	\begin{lemma}
		\label{lem:static random walks congestion}
		Let $E_t$ denote the set of new edges of a hypercube at time $t$. If all edges in $E_t$ are sampled at time $t$, then the congestion in $G_t$ induced by the legs of the sampled paths is $O(\tmix\cdot \log^2n)$ with high probability. 
	\end{lemma}
	\begin{proof}
		Let $k := 2|E_t|$ denote the number of legs we need to sample. Lemma \ref{lem:path success} tells us that with high probability, each leg from edge $e$ can be chosen out of $O(\log n)$ independent random walks from $e$. Therefore, with high probability, the congestion of all legs is no more than a congestion of $k\cdot O(\log n)$ random walks. Since each edge is a source of at most $O(\log n)$ legs, each edge is a source of $O(\log^2 n)$ random walks. Further, each random walk is of length $\tmix$. The standard random walk congestion argument then tells us that the congestion of these random walks is at most $O(\tmix \cdot \log^2 n)$ whp.
	\end{proof}
	
	Combining Lemma \ref{lem:static random walks congestion} and Theorem \ref{thm:greedy temporal aggregation main}, we can prove Lemma \ref{lem:rw congestion}.
	\rwCongestion*
	\begin{proof}
		Invoke Theorem \ref{thm:greedy temporal aggregation main} with a static guarantee of Lemma \ref{lem:static random walks congestion} with a load being a congestion of edge $e$.
	\end{proof}

	The next lemma summarizes the analysis and states that the total congestion is small.
	\lowCongest*
	\begin{proof}
		We split the congestion of a given edge $e \in E(G)$ into two parts: congestion from legs and congestion from greedy paths' projections. The congestion from legs is $\widetilde{O}(1)$ with high probability by Lemma \ref{lem:rw congestion}, the congestion from greedy paths' projections is $\widetilde{O}(1)$ by Lemma \ref{lem:greedy per edge}.
	\end{proof}

	\subsubsection{Computation}
	This subsection bounds the total computation performed by an algorithm by the polylog times the number of edge deletions performed by the adversary.
	
	Lemma \ref{lem:path success} states that the algorithm succeeds in establishing a path in a few tries. What is left to show is the bound on the total number of times the algorithm will attempt to sample a path. 
	
	\begin{lemma}
		\label{lem:few path sampling attempts}
		If the adversary performs $q$ deletions in total, w.h.p., the number of attempts by the algorithm to sample a path will be $\widetilde{O}(q)$.
	\end{lemma}
	\begin{proof}
		The algorithm attempts to sample a path for three reasons: (1) It establishes an edge, (2) It fixes an edge, and (3) It resamples an edge due to temporal aggregation. 
		
		Denote the total number of newly established, fixed, and temporarily aggregated edges with $X$, $Y$, and $Z$, respectively.
		
		The number of new edges established per deletion is $O(\log n)$, hence $X = O(q\cdot \log n)$. According to Lemma \ref{lem:low congest}, w.h.p., the number of paths that need to be fixed after a single edge deletion is $\widetilde{O}(1)$, hence $Y= \widetilde{O}(q)$. Finally, according to Theorem \ref{thm:greedy temporal aggregation main}, $Z = O(\log n \cdot (X + Y)) = \widetilde{O}(q)$. 
	\end{proof}
	
	We therefore get the following total computation guarantee.
	
	\begin{lemma}
		\label{lem:low computation embedding}
		If the adversary performs $q$ deletions in total, w.h.p., the total computation performed by the algorithm is $\widetilde{O}(q)$.
	\end{lemma}
	\begin{proof}
		By Lemma \ref{lem:few path sampling attempts} we know that whp the number of path resampling attempts is $\widetilde{O}(q)$. Lemma \ref{lem:path success} stipulates that one attempt requires $O(\log n)$ tries with high probability.
		Therefore, whp, the total computation is $\widetilde{O}(q)$.
	\end{proof}
	
	Finally, we prove Theorem \ref{thm:hypercube embedding}. 
	
	\hypercubeEmbeddingThm*
	\begin{proof}
		Apply Lemmas \ref{lem: hypercube embedding dilation}, \ref{lem:low congest}, \ref{lem:low computation embedding}.
	\end{proof}

	\section{Dynamic Hypercube Routing}
	This section discusses routing demands in a dynamic generalized hypercube---the building block we use in our general dynamic expander routing algorithm when provided an embedding. 
	
	\subsection{Problem Statement}
	Initially, the algorithm is given a generalized hypercube $H_1$ and a set of demands $\mathcal{D}_1 \subseteq V(H_1)\times V(H_1)$. The algorithm may perform a polynomial-time setup phase, after which it outputs a collection of paths $\mathcal{P}_1$ in $H_1$ such that every demand in $\mathcal{D}_1$ is served by some path in $\mathcal{P}_1$.
	
	At each time step $t>1$, an adaptive adversary---having observed the current maintained paths---modifies the instance. Specifically, it deletes some nodes and edges from $H_{t-1}$ and inserts some edges to obtain a new $H_t$. Also, it may add or delete demands to obtain a new demand set $\mathcal{D}_t$ from $\mathcal{D}_{t-1}$. After these changes, the algorithm must update its solution from $\mathcal{P}_{t-1}$ to $\mathcal{P}_t$ so that every demand in $\mathcal{D}_t$ is served.
	
	We impose two constraints on the adversary. First, we require that every $H_t$ is a generalized hypercube. Second, for every time $t$ and every node $u \in V(H_t)$, $u$ is a source or a destination of at most $1$ demand in $\mathcal{D}_t$.
	
	The goal is to maintain a set of paths $\mathcal{P}_t$ with small maximum dilation and small maximum congestion at all times, while also minimizing the computation performed per adversarial modification (vertex deletions, edge insertions/deletions, and demand insertions/deletions).

	\subsection{Static Routing in Generalized Hypercube}
	\label{sec:static routing}
	We now give a Valiant-like oblivious routing procedure that performs well in static generalized hypercubes.
	
	\begin{algorithm}
		\caption{Oblivious routing from $s$ to $t$ in generalized hypercube.}
		\label{alg:static routing}
		\begin{algorithmic}
			\State $w \gets$ uniformly random node
			\State $p_s \gets$ greedy path from $s$ to $w$
			\State $p_t \gets$ greedy path from $w$ to $t$
			\State \textbf{Output} concatenation of $p_s$ and $p_t$
		\end{algorithmic}
	\end{algorithm}
	
	This procedure results in low congestion with high probability.
	
	\begin{lemma}
		\label{lem:static hypercube routing}
		Consider a hypercube $H$ on $n$ nodes and a set of demands $\mathcal{D}$ such that each node in $H$ is a source or a destination of at most one demand. Assume we route all demands in $\mathcal{D}$ simultaneously and independently by the Algorithm \ref{alg:static routing}. Then, with high probability, the maximum congestion induced by the paths is $O(\log n)$.
	\end{lemma}
	\begin{proof}
		Fix an edge $e\in E(H)$. We bound the number of paths that traverse $e$; a union bound over all edges then yields the maximum congestion.
		
		Consider only the $s\to w$ segments (the $w\to t$ segments are identical by symmetry since $w$ is sampled uniformly and independently for each demand). For each demand $(s,t)\in\mathcal D$ define
		\[
		X_{s,e}=\mathbf 1[\text{the greedy path from $s$ to its sampled $w$ uses $e$}],
		\qquad
		X_e=\sum_{(s,\cdot)\in\mathcal D} X_{s,e}.
		\]
		We show $\mathbb E[X_e]\le 2$ and then apply Chernoff.
		
		Write $e=(a,b)$ with $a<b$ and let its \emph{stretch} be $\ell=b-a=2^i$. If $s\in(a,b)$ then a greedy path from $s$ cannot traverse $(a,b)$, so $\Pr[X_{s,e}=1]=0$.
		
		Now fix $s\le a$. If the greedy path uses $(a,b)$, then upon reaching $a$ the next step must be exactly $\ell$, which by the greedy rule happens iff the remaining distance to $w$ satisfies $w-a\in[\ell,2\ell)$, i.e.\ $w\in[b,b+\ell)$. Hence for every $s\le a$,
		\[
		\Pr[X_{s,e}=1]\le \Pr[w\in[b,b+\ell)]=\ell/n.
		\]
		Moreover, only a few sources $s\le a$ can reach $a$ and then take an $\ell$-jump: along a greedy path, step lengths are strictly decreasing powers of two, so before taking the $\ell$-edge, all earlier steps must have lengths in
		$\{2^{i+1},2^{i+2},\dots,2^{\lfloor\log_2 n\rfloor}\}$.
		Choosing any subset of these lengths uniquely determines the total displacement from $s$ to $a$, hence yields at most one such source $s$. Therefore the number of sources $s\le a$ whose greedy path \emph{can} traverse $e$ is at most
		\[
		2^{\lfloor\log_2 n\rfloor-i}\le n/2^i = n/\ell.
		\]
		Since $\mathcal D$ contains at most one demand per source, we get
		\[
		\sum_{\substack{(s,\cdot)\in\mathcal D\\ s\le a}} \mathbb E[X_{s,e}]
		\;\le\;
		\frac{n}{\ell}\cdot \frac{\ell}{n}
		\;=\;1.
		\]
		By symmetry (reversing the order on $[n]$), the same bound holds for sources $s\ge b$, contributing at most $1$ in expectation. Thus
		\[
		\mathbb E[X_e]=\sum_{(s,\cdot)\in\mathcal D}\mathbb E[X_{s,e}] \le 2.
		\]
		
		For fixed $e$, the variables $\{X_{s,e}\}_{(s,\cdot)\in\mathcal D}$ are independent because the intermediate vertices $w$ are sampled independently across demands. Hence, by a Chernoff bound, there is an absolute constant $c>0$ such that
		\[
		\Pr[X_e\ge c\log n]\le n^{-3}.
		\]
		The generalized hypercube has $O(n\log n)$ edges, so a union bound implies
		\[
		\Pr\Big[\exists e\in E(H): X_e\ge c\log n\Big]
		\;\le\; O(n\log n)\cdot n^{-3}=o(1).
		\]
		Therefore, with high probability, every edge has congestion $O(\log n)$ from the $s\to w$ segments. The same holds for the $w\to t$ segments, and thus the total congestion per edge is $O(\log n)$ with high probability.
	\end{proof}
	
	We also observe from the description of the algorithm that it takes little computation.
	
	\begin{lemma}
		\label{lem:static routing computation}
		The Algorithm \ref{alg:static routing} requires $O(\log n)$ computation.
	\end{lemma}
	
	We remark that the Algorithm \ref{alg:static routing} can be used in the precomputation phase of the Dynamic Generalized Hypercube Routing problem, yielding $O(\log n)$ maximal congestion whp and $O(n\log n)$ precomputing time. We now explain how we maintain this routing.
	
	\subsection{Dynamic Algorithm}
	To lift the static routing scheme to the dynamic setting with an adaptive adversary, we use Greedy Temporal Aggregation of \cite{temporalAggregation}.
	
	\begin{algorithm}
		\caption{Dynamic Hypercube Routing}
		\label{alg:dynamic hypercube routing}
		
		We map our setting to the framework of~\cite{temporalAggregation}. Each demand is an object. We say an object is adversarially added whenever the adversary introduces a new demand, or adversarially resampled if the adversary deletes an edge that the demand's current route uses. Resampling an object corresponds to recomputing its route using the Algorithm \ref{alg:static routing}. With this, we apply Greedy Temporal Aggregation.
	\end{algorithm}
	
	\begin{theorem}
		\label{thm:dynamic hypercube routing}
		Consider a sequence of graphs $G_1, \ldots, G_T$ on $n$ vertices with $T = \mathrm{poly}(n)$, such that each $G_t$ is a generalized hypercube. Consider also a sequence of demands $\mathcal{D}_1, \ldots, \mathcal{D}_T$, such that for every $t \in [T]$ and $v \in V(G_t)$, $v$ is a source or a destination of at most $1$ demand in $\mathcal{D}_t$. 
		
		There exists an algorithm that maintains routing paths for $\mathcal{D}_t$ against an adaptive adversary such that the congestion stays $O(\log^2 n)$ at all times $t\in[T]$ whp. Moreover, if the adversary performs a total of $q$ edge deletions and demand insertions, the algorithm performs $O(q \cdot \log^4 n)$ computation whp.
	\end{theorem}
	\begin{proof}
		Algorithm \ref{alg:dynamic hypercube routing} instantiates Corollary \ref{cor:temporal aggregation 4 routing} with a static routing Algorithm \ref{alg:static routing}. Lemmas \ref{lem:static hypercube routing} and \ref{lem:static routing computation} then imply the statement of the theorem. 
	\end{proof}
	
	\section{Putting it Together}
	In this section, we combine our Dynamic Hypercube Embedding and Dynamic Hypercube Routing algorithm to obtain a solution for the original Dynamic Expander Routing Problem. 
	
	\subsection{Algorithm for Dynamic Expander Routing}
	\begin{algorithm}
		\caption{Dynamic Expander Routing}
		\label{alg:compile}
		First, in the setup phase, we do an embedding $\xi$ of a Generalized Hypercube $H_1$ onto $G_1$, e.g., according to the algorithm presented in \cite{chang2024deterministic}. Throughout the process, we maintain this embedding according to Algorithm \ref{alg:dynamic hypercube embedding}.\\ 
		
		The initial set of demands for $G_1$ is served as follows. For each initial demand $(u,v)$, we assign it to a spare edge $e_u$ of $u$ and a spare edge $e_v$ of $v$. Then, we serve the demand $(\xi^{-1}(e_u), \xi^{-1}(e_v))$ in $H$ by the Algorithm \ref{alg:static routing}. Finally, we serve $(u, v)$ in $G$ by the projection of a path serving $(\xi^{-1}(e_u), \xi^{-1}(e_v))$ in $H$.\\
		
		We will be maintaining two sets of demands---one, the set on the underlying graph $G$ called $\mathcal{D}^G$, and one on the embedded hypercube $H$ called $\mathcal{D}^H$. We handle serving demands from $\mathcal{D}^H$ by Algorithm \ref{alg:dynamic hypercube routing}. We now focus on handling demands from $\mathcal{D}^G$, given that the demands in $\mathcal{D}^H$ are handled. When an adversary adds or deletes a demand for $G$ at time $t$, we add or delete it respectively in $\mathcal{D}^G_t$. Also, if an adversary deletes an edge in $G_{t-1}$ through which a path serving some demand $d$ was passing, we delete and add $d$ back to $\mathcal{D}^G_t$ explicitly.\\
		
		The added demand $(u, v) \in \mathcal{D}_t^G$ is served similarly to the initial demands. First, we assign it to a spare edge $e_u$ of $u$ and a spare edge $e_v$ of $v$. Then, we add $(\xi^{-1}(e_u), \xi^{-1}(e_v))$ to $\mathcal{D}_t^H$. Once the demand $(\xi^{-1}(e_u), \xi^{-1}(e_v))$ is served in $H$ by Algorithm \ref{alg:dynamic hypercube routing} via a path $p$, we serve it in $G$ by the projection of $p$.\\ 
	\end{algorithm}

	\subsection{Analysis}
	We now focus on analyzing the correctness of the Algorithm \ref{alg:compile}, as well as the dilation and congestion induced by it at every time, and the total computation. 
	
	\begin{lemma}[Correctness]
		\label{lem:compiled correctness}
		The Algorithm \ref{alg:compile} fulfills the assumptions of the subroutines it is using and correctly solves the Expander Routing Problem.
	\end{lemma}
	\begin{proof}
		We can use a Dynamic Hypercube Embedding Algorithm \ref{alg:dynamic hypercube embedding} since the graph $G$ always satisfies the low mixing time Assumption \ref{assump:always expander}.
		
		Note that for an added demand $(u, v)$ by the vertex degree feasibility Assumption \ref{assump:vertex feasibility}, there will always be a spare edge $e_v$ and $e_u$. Furthermore, note that since an edge in $G$ is assigned at most one demand, a vertex in $H$ is as well, so the condition of Theorem \ref{thm:dynamic hypercube routing} is fulfilled.
		
		Lastly, we observe that the paths maintained by the algorithm at time $t$ indeed serve $\mathcal{D}_t$ since the projection of a continuous path in $H$ is a continuous path in $G$ and it starts and ends at the edges incident to the endpoints of the demand. 
	\end{proof}

	\begin{lemma}[Dilation]
		\label{lem:compiled dilation}
		The dilation induced by the Algorithm \ref{alg:compile} is $\widetilde{O}(1)$.
	\end{lemma}
	\begin{proof}
		The dilation of a path between $(u, v) \in V^2(G)$ is at most the number of edges in the hypercube path serving $(\xi^{-1}(e_u), \xi^{-1}(e_v))$ times the maximal length of the path corresponding to a hypercube edge. The prior is the length of two greedy paths, which is bounded by Lemma \ref{lem:greedy length} by $O(\log n)$. And the latter is bounded by Lemma \ref{lem: hypercube embedding dilation} by $\widetilde{O}(1)$. Hence the dilation of a path serving $(u, v)$ is $\widetilde{O}(1)$.
	\end{proof}

	\begin{lemma}[Congestion]
		\label{lem:compiled congestion}
		The congestion induced by the Algorithm \ref{alg:compile} is $\widetilde{O}(1)$ with high probability.
	\end{lemma}
	\begin{proof}
		The congestion of an edge $e\in E(G)$ is at most the maximal number of hypercube edge projections crossing $e$ times the maximal congestion in the hypercube. The prior is bounded whp by $\widetilde{O}(1)$ by Lemma \ref{lem:low congest} and the latter is bounded by $O(\log^2n)$ whp by Theorem \ref{thm:dynamic hypercube routing}.
	\end{proof}
	
	\begin{lemma}[Computation]
		\label{lem:compiled computation}
		If the adversary performs $q$ edge deletions and demand insertions in total, the computation of the Algorithm \ref{alg:compile} is $\widetilde{O}(q)$ with high probability.
	\end{lemma}
	\begin{proof}
		We start by observing that the setup phase indeed requires polynomial time. In particular, the algorithm for the initial hypercube embedding of \cite{chang2024deterministic} has polynomial running time. And the initial demand serving is $O(m\log n)$ computation, since there are at most $O(m)$ demands according to Assumption \ref{assump:vertex feasibility} and the Algorithm \ref{alg:static routing} runs in $O(\log n)$ computation per demand according to Lemma \ref{lem:static routing computation}.
		
		After the setup phase, we have two sources of computation: maintaining the hypercube embedding and maintaining the demands in the hypercube. The first one is guaranteed to result in $\widetilde{O}(q)$ total computation with high probability by Lemma \ref{lem:low computation embedding}. To apply Theorem \ref{thm:dynamic hypercube routing} for the second one, we first need to bound the total number of deleted edges and added demands in the hypercube. The deleted edges are straightforward: the edge is only deleted when a node is deleted, and a node in a hypercube is deleted when the corresponding edge in $G$ is deleted. Once a node in the hypercube is deleted, all the incident edges are deleted, and there are $O(\log n)$ of those. Hence, if $q$ edges are deleted in $G$, $O(q\cdot \log n)$ edges are deleted in $H$. The demand insertions are for two reasons: either it is a completely new demand inserted, or it is a reinserted demand for which an edge in $G$ on a path serving this demand was deleted. By definition, there can be no more than $q$ newly inserted demands. As for reinserted demands, the Lemma \ref{lem:compiled congestion} stipulates that whp there are at most $\widetilde{O}(1)$ paths that are using an edge, hence per $q$ edge deletions, at most $\widetilde{O}(q)$ demands will be reinserted. This allows us to conclude that if an adversary does $q$ demand insertions and edge deletions in $G$, it does $q' = \widetilde{O}(q)$ demand insertions and edge deletions in $H$. Therefore, Theorem \ref{thm:dynamic hypercube routing} gives us that the total computation for maintaining demands in the hypercube is $O(q'\cdot \log^4 n) = \widetilde{O}(q)$.
		
	\end{proof}
	
	Finally, we prove our main theorem.

	\dynamicExpanderRoutingThm*
	\begin{proof}
		The Algorithm \ref{alg:compile} fulfills the Theorem by Lemmas \ref{lem:compiled correctness}, \ref{lem:compiled dilation}, \ref{lem:compiled congestion}, \ref{lem:compiled computation}.
	\end{proof}

	\bibliographystyle{alpha}
	\bibliography{refs}
	
	\appendix
	\section{Extending Temporal Aggregation Framework}
	\label{sec:appendix:extending ta}
	In this section, we show that Theorem \ref{thm:greedy temporal aggregation main} holds as well if one is allowed to add and delete objects from the system. The only part of the original proof to be updated is the one concerning a few resamples done by the algorithm.
	
	\begin{restatable}{theorem}{GreedyTemporalAggregationPlusThm}
		\label{thm:greedy temporal aggregation plus}
		If the system evolves for $T$ time steps, and in each time step a static realization has a load at most $L$ with probability at least $1 - \varepsilon$, then Greedy Temporal Aggregation ensures that at time $T$ the joint realization has a load at most $L\cdot O(\log n)$ with probability at least $1 - \varepsilon \cdot T$.
		
		Moreover, if the adversary does $q$ samples and object insertions by the time $T$, Greedy Temporal Aggregation performs no more than $q\cdot O(\log n)$ samples.
	\end{restatable}
	\begin{proof}[Proof of few resamples.]
		We assume that the initial sample is at time $0$.
		
		To prove that the algorithm does a bounded number of additional samples, we conduct an amortized analysis. We say that each object has coins in its wallet and in its bequest savings. We also say that objects at time $0$ have a bank. 
		
		When the adversary samples or adds an object, we assume this object receives $\log_{3/2}n$ coins to its wallet and $4$ coins to its bequest from the adversary. Additionally, when the adversary samples or adds an object, it adds $3/2$ coins to the bank. Also, if the adversary samples an object that was previously sampled at time $t$ and there were $k$ objects sampled at time $t$, they all receive $4/k$ coins to their wallet from the adversary. This way, if an adversary samples or adds an object, it spends at most $\log_{3/2}n + 9\frac{1}{2}$ coins.
		
		Further, if an adversary deletes an object that was previously sampled at time $t > 0$ and there were $k$ objects sampled at time $t$, they all receive $4/k$ coins to their wallet from the bequest of the deleted object.
		
		When an algorithm samples an object sampled at time $t$, it takes one coin from its wallet if $t > 0$ or one coin from the bank if $t = 0$.
		
		Denote with $group(u)$ a set of objects sampled at the same time step as $u$ and with $wallet(u)$ the number of coins $u$ has in its wallet. Consider the potential function of $u$: $\Phi(u) = \log_{3/2}(|group(u)|) + wallet(u)$. We claim that this potential is monotonically increasing for all objects once they are sampled at time $t > 0$. Indeed, consider the cases when this potential is changed. 
		\begin{itemize}
			\item The object $u$ is added to the group. Then all the potentials only increase.
			\item The object $u$ is sampled by the adversary. Then, the $|group(u)|$ may decrease (at most) from $n$ to $1$, hence the $\log_{3/2}(|group(u)|)$ from $\log_{3/2}(n)$ to $0$, but adversary gives the object $\log_{3/2}(n)$ coins to its wallet, so $\Phi(u)$ can not decrease.
			\item Some other object from $group(u)$ is sampled by the adversary. Assume that before the sample $|group(u)| = k \geq 2$. Then $\Delta\Phi(u) = \log_{3/2}(k - 1) - \log_{3/2}(k) + 4/k > 0$.
			\item Some other object from $group(u)$ is deleted by the adversary. Assume that before the deletion $|group(u)| = k \geq 2$. Then, as in the previous case, $\Delta\Phi(u) = \log_{3/2}(k - 1) - \log_{3/2}(k) + 4/k > 0$.
			\item All the objects in $group(u)$ are sampled by the algorithm. This means that they are being sampled together with some other group, whose size is at least $\frac{1}{2}|group(u)|$. Therefore, the new size $|group'(u)| \geq |group(u)| + \frac{1}{2}|group(u)| = \frac{3}{2}|group(u)|$ and hence $\Delta\Phi(u) = \log_{3/2}(|group'(u)|) - \log_{3/2}(|group(u)|) - 1 \geq 0$.
		\end{itemize}
		
		We note that if an object is deleted at time $t > 0$, it was already either sampled or added, hence it has at least $4$ coins in its bequest and can pay to the nodes of its group.
		
		We can now prove that the algorithm will always be able to take a coin. First, consider sampling an object $u$ that was last sampled at time $t > 0$. Indeed, we've shown that $\Phi(u) \geq \log_{3/2}n$. If the object is sampled, it means its group's size is increased by at least $3/2$ times, hence it was less than $\frac{n}{3/2}$, hence the number of coins in the wallet of an object was at least $1$, since $\log_{3/2}(|group(u)|) + coins(u) \geq \log_{3/2}n$ and $\log_{3/2}(|group(u)|) \leq \log_{3/2}n - 1$.
		
		Next, consider an algorithm samples an object that was last sampled at time $0$. Denote the group size of objects at time $0$ with $n_0$, the size of some other group at time $t > 0$ with which it gets merged with $n_t$. By the definition of the algorithm, we know that $n_0 \leq 3/2 n_t$. However, for every object at time $t > 0$, the adversary gave to a bank $3/2$ coins, hence there are at least $3/2n_t$ coins in the bank, and we can use those to pay for sampling objects at time $0$.
		
		Finally, we conclude that 
		\begin{align*}
			\text{Total \# of samples by the alg} &\leq \text{\# of coins introduced to the system}\\
			&\leq \text{\# of adversarial samples and object additions}\cdot(\log_{3/2}n + 9\frac{1}{2}).
		\end{align*}
	\end{proof}
	
\end{document}

%% file: pics/hypercube_embedding1.tikz.tex
\begin{tikzpicture}[>=Stealth, line cap=round, line join=round, font=\large]

\tikzset{
  pt/.style={circle, fill=black, inner sep=1.8pt},
  lift/.style={gray!70, dashed, line width=0.9pt},
  arr/.style={gray!70, dashed, -{Stealth[length=3.0mm,width=2.2mm]}, line width=0.9pt},
  bluepath/.style={blue!70, line width=1.1pt},
}

\coordinate (HTL) at (-5.2, 3.0);
\coordinate (HTR) at ( 5.2, 3.0);
\coordinate (HBL) at (-6.3, 1.4);
\coordinate (HBR) at ( 6.3, 1.4);
\draw[line width=1.0pt] (HBL) -- (HTL) -- (HTR) -- (HBR) -- cycle;

\node[anchor=east, font=\Huge] at (-7.2, 2.2) {H};

\coordinate (h1) at (-3.9, 2.05);
\coordinate (h2) at ( 3.9, 2.05);
\node[pt] at (h1) {};
\node[pt] at (h2) {};

\foreach \p in {h1,h2} {
  \draw[gray!70, line width=0.9pt] ($(\p)+(0,-0.03)$) -- ($(\p)+(-0.12,-0.25)$);
  \draw[gray!70, line width=0.9pt] ($(\p)+(0,-0.03)$) -- ($(\p)+( 0.00,-0.27)$);
  \draw[gray!70, line width=0.9pt] ($(\p)+(0,-0.03)$) -- ($(\p)+( 0.12,-0.25)$);
}

\coordinate (hmid) at (0, 2.55);
\draw[bluepath] (h1) -- (hmid) -- (h2);

\coordinate (Gcenter) at (0, -2.0);
\draw[line width=1.0pt] (Gcenter) ellipse (6.0 and 1.6);

\node[anchor=east, font=\Huge] at (-7.2, -2.2) {G};

\coordinate (u) at (-3.4, -2.55);
\coordinate (v) at ( 3.4, -2.55);
\coordinate (u2) at ($(u)+(-0.95,0.55)$);
\coordinate (v2) at ($(v)+( 0.95,0.55)$);

\draw[line width=1.0pt] (u2) -- (u);
\draw[line width=1.0pt] (v2) -- (v);

\node[pt] at (u) {};
\node[pt] at (v) {};
\node[anchor=west] at ($(u)+(0.12,-0.05)$) {$\mathbf{s}$};
\node[anchor=west] at ($(v)+(-0.65,-0.05)$) {$\mathbf{t}$};

\coordinate (m1) at ($(u2)!0.5!(u)$);
\coordinate (m2) at ($(v2)!0.5!(v)$);

\node[anchor=north east] at ($(m1)+(0.15,0)$) {$e_1$};
\node[anchor=north east] at ($(m2)+(0.6,0)$) {$e_2$};

\coordinate (m1v) at ($(h1 |- m1)$); %
\coordinate (m2v) at ($(h2 |- m2)$); %

\draw[lift] (h1) -- (m1v);
\draw[lift] (h2) -- (m2v);

\draw[arr] (h1) -- (m1v);
\draw[arr] (h2) -- (m2v);

\node at ($(h1)!0.55!(m1v) + (0.35,0)$) {$\xi$};
\node at ($(h2)!0.55!(m2v) + (0.35,0)$) {$\xi$};

\end{tikzpicture}

%% file: pics/hypercube_embedding2.tikz.tex
\begin{tikzpicture}[>=Stealth, line cap=round, line join=round, font=\large]

\tikzset{
  pt/.style={circle, fill=black, inner sep=1.8pt},
  lift/.style={gray!70, dashed, line width=0.9pt},
  arr/.style={gray!70, dashed, -{Stealth[length=3.0mm,width=2.2mm]}, line width=0.9pt},
  bluepath/.style={blue!70, line width=1.1pt},
}

\coordinate (HTL) at (-5.2, 3.0);
\coordinate (HTR) at ( 5.2, 3.0);
\coordinate (HBL) at (-6.3, 1.4);
\coordinate (HBR) at ( 6.3, 1.4);
\draw[line width=1.0pt] (HBL) -- (HTL) -- (HTR) -- (HBR) -- cycle;

\coordinate (h1) at (-3.9, 2.05);
\coordinate (h2) at ( 3.9, 2.05);
\node[pt] at (h1) {};
\node[pt] at (h2) {};

\foreach \p in {h1,h2} {
  \draw[gray!70, line width=0.9pt] ($(\p)+(0,-0.03)$) -- ($(\p)+(-0.12,-0.25)$);
  \draw[gray!70, line width=0.9pt] ($(\p)+(0,-0.03)$) -- ($(\p)+( 0.00,-0.27)$);
  \draw[gray!70, line width=0.9pt] ($(\p)+(0,-0.03)$) -- ($(\p)+( 0.12,-0.25)$);
}

\coordinate (hmid) at (0, 2.55);
\draw[bluepath] (h1) -- (hmid) -- (h2);

\coordinate (Gcenter) at (0, -2.0);
\draw[line width=1.0pt] (Gcenter) ellipse (6.0 and 1.6);

\coordinate (u) at (-3.4, -2.55);
\coordinate (v) at ( 3.4, -2.55);
\coordinate (u2) at ($(u)+(-0.95,0.55)$);
\coordinate (v2) at ($(v)+( 0.95,0.55)$);

\draw[line width=1.0pt] (u2) -- (u);
\draw[line width=1.0pt] (v2) -- (v);

\node[pt] at (u) {};
\node[pt] at (v) {};
\node[anchor=west] at ($(u)+(0.12,-0.05)$) {$\mathbf{s}$};
\node[anchor=west] at ($(v)+(-0.65,-0.05)$) {$\mathbf{t}$};

\coordinate (m1) at ($(u2)!0.5!(u)$);
\coordinate (m2) at ($(v2)!0.5!(v)$);

\node[anchor=north east] at ($(m1)+(0.15,0)$) {$e_1$};
\node[anchor=north east] at ($(m2)+(0.6,0)$) {$e_2$};

\coordinate (wL) at ($(u2)$);
\coordinate (wR) at ($(v2)$);

\draw[bluepath]
  (wL)
  .. controls (-2.7, -1.25) and (-2.2,-1.55) .. (-1.6,-2.05)
  .. controls (-1.0,-2.55) and (-0.4,-1.45) .. ( 0.2,-1.95)
  .. controls ( 0.8,-2.45) and ( 1.5,-1.55) .. ( 2.2,-2.05)
  .. controls ( 2.8,-2.55) and ( 3.2,-2.10) .. (wR);

\coordinate (m1v) at ($(h1 |- m1)$); %
\coordinate (m2v) at ($(h2 |- m2)$); %

\draw[lift] (h1) -- (m1v);
\draw[lift] (h2) -- (m2v);

\draw[arr] (h1) -- (m1v);
\draw[arr] (h2) -- (m2v);

\node at (-0.3,0.2) {$\xi$};

\foreach \x in {-2.0,0.0,2.0} {
  \draw[lift] (\x,2.10) -- (\x,-1.05);
  \draw[arr]  (\x,-1.05) -- (\x,-1.70);
}

\foreach \x in {-2.0,0.0,2.0} {
  \draw[arr] (\x,1.78) -- (\x,2.22);
}

\end{tikzpicture}

%% file: main.bbl
\begin{thebibliography}{CHLP21}

\bibitem[BMI10]{busch2010optimal}
Costas Busch and Malik Magdon-Ismail.
\newblock Optimal oblivious routing in hole-free networks.
\newblock In {\em International Conference on Heterogeneous Networking for Quality, Reliability, Security and Robustness}, pages 421--437. Springer, 2010.

\bibitem[BMIX05]{busch2005oblivious}
Costas Busch, Malik Magdon-Ismail, and Jing Xi.
\newblock Oblivious routing on geometric networks.
\newblock In {\em Proceedings of the seventeenth annual ACM symposium on Parallelism in algorithms and architectures}, pages 316--324, 2005.

\bibitem[BMIX08]{busch2008optimal}
Costas Busch, Malik Magdon-Ismail, and Jing Xi.
\newblock Optimal oblivious path selection on the mesh.
\newblock {\em IEEE Transactions on Computers}, 57(5):660--671, 2008.

\bibitem[CHLP21]{censor2021sparsity}
Keren Censor-Hillel, Dean Leitersdorf, and Volodymyr Polosukhin.
\newblock On sparsity awareness in distributed computations.
\newblock In {\em Proceedings of the 33rd ACM Symposium on Parallelism in Algorithms and Architectures}, pages 151--161, 2021.

\bibitem[CHS24]{chang2024deterministic}
Yi-Jun Chang, Shang-En Huang, and Hsin-Hao Su.
\newblock Deterministic expander routing: Faster and more versatile.
\newblock In {\em Proceedings of the 43rd ACM Symposium on Principles of Distributed Computing}, pages 194--204, 2024.

\bibitem[CPP20]{chatterjee2020distributed}
Soumyottam Chatterjee, Gopal Pandurangan, and Nguyen~Dinh Pham.
\newblock Distributed mst: A smoothed analysis.
\newblock In {\em Proceedings of the 21st International Conference on Distributed Computing and Networking}, pages 1--10, 2020.

\bibitem[CPZ19]{trianglesThroughDecomposition}
Yi-Jun Chang, Seth Pettie, and Hengjie Zhang.
\newblock Distributed triangle detection via expander decomposition.
\newblock In {\em Proceedings of the Thirtieth Annual ACM-SIAM Symposium on Discrete Algorithms}, pages 821--840. SIAM, 2019.

\bibitem[CS19]{trianglesThroughDecomposition2}
Yi-Jun Chang and Thatchaphol Saranurak.
\newblock Improved distributed expander decomposition and nearly optimal triangle enumeration.
\newblock In {\em Proceedings of the 2019 ACM Symposium on Principles of Distributed Computing}, pages 66--73, 2019.

\bibitem[CS20]{deterministicExpanderDecomposition}
Yi-Jun Chang and Thatchaphol Saranurak.
\newblock Deterministic distributed expander decomposition and routing with applications in distributed derandomization.
\newblock In {\em 2020 IEEE 61st Annual Symposium on Foundations of Computer Science (FOCS)}, pages 377--388. IEEE, 2020.

\bibitem[GHZ21]{ghaffari2021hop}
Mohsen Ghaffari, Bernhard Haeupler, and Goran Zuzic.
\newblock Hop-constrained oblivious routing.
\newblock In {\em Proceedings of the 53rd Annual ACM SIGACT Symposium on Theory of Computing}, pages 1208--1220, 2021.

\bibitem[GKS17]{originalExpanderRouting}
Mohsen Ghaffari, Fabian Kuhn, and Hsin-Hao Su.
\newblock Distributed mst and routing in almost mixing time.
\newblock In {\em Proceedings of the ACM Symposium on Principles of Distributed Computing}, pages 131--140, 2017.

\bibitem[GL18]{newExpanderRouting}
Mohsen Ghaffari and Jason Li.
\newblock New distributed algorithms in almost mixing time via transformations from parallel algorithms.
\newblock {\em arXiv preprint arXiv:1805.04764}, 2018.

\bibitem[HP26]{temporalAggregation}
Bernhard Haeupler and Anton Paramonov.
\newblock Maintaining random assignments under adversarial dynamics.
\newblock {\em arXiv preprint arXiv:2604.05606}, 2026.

\bibitem[HRG22]{haeupler2022hop}
Bernhard Haeupler, Harald R{\"a}cke, and Mohsen Ghaffari.
\newblock Hop-constrained expander decompositions, oblivious routing, and distributed universal optimality.
\newblock In {\em Proceedings of the 54th Annual ACM SIGACT Symposium on Theory of Computing}, pages 1325--1338, 2022.

\bibitem[Rab89]{rabin1989efficient}
Michael~O Rabin.
\newblock Efficient dispersal of information for security, load balancing, and fault tolerance.
\newblock {\em Journal of the ACM (JACM)}, 36(2):335--348, 1989.

\bibitem[Rac02]{racke2002minimizing}
Harald Racke.
\newblock Minimizing congestion in general networks.
\newblock In {\em The 43rd Annual IEEE Symposium on Foundations of Computer Science, 2002. Proceedings.}, pages 43--52. IEEE, 2002.

\bibitem[R{\"a}c08]{racke2008optimal}
Harald R{\"a}cke.
\newblock Optimal hierarchical decompositions for congestion minimization in networks.
\newblock In {\em Proceedings of the fortieth annual ACM symposium on Theory of computing}, pages 255--264, 2008.

\bibitem[R{\"a}c09]{racke2009survey}
Harald R{\"a}cke.
\newblock Survey on oblivious routing strategies.
\newblock In {\em Conference on Computability in Europe}, pages 419--429. Springer, 2009.

\bibitem[Sch06]{scheideler2006universal}
Christian Scheideler.
\newblock {\em Universal routing strategies for interconnection networks}, volume 1390.
\newblock Springer, 2006.

\bibitem[SV19]{su2019distributed}
Hsin-Hao Su and Hoa~T Vu.
\newblock Distributed data summarization in well-connected networks.
\newblock {\em arXiv preprint arXiv:1908.00236}, 2019.

\bibitem[SW19]{pruning}
Thatchaphol Saranurak and Di~Wang.
\newblock Expander decomposition and pruning: Faster, stronger, and simpler.
\newblock In {\em Proceedings of the Thirtieth Annual ACM-SIAM Symposium on Discrete Algorithms}, pages 2616--2635. SIAM, 2019.

\bibitem[Upf84]{upfal1984efficient}
Eli Upfal.
\newblock Efficient schemes for parallel communication.
\newblock {\em Journal of the ACM (JACM)}, 31(3):507--517, 1984.

\bibitem[VB81]{valiant1981universal}
Leslie~G Valiant and Gordon~J Brebner.
\newblock Universal schemes for parallel communication.
\newblock In {\em Proceedings of the thirteenth annual ACM symposium on Theory of computing}, pages 263--277, 1981.

\end{thebibliography}
